# SOCIO-DEMOGRAPHIC GOALS WITHIN DIGITALIZATION ENVIRONMENT: A GENDER ASPECT[1]

**ZOLOTAREVA Olga A.[1],**

**BEZRUKOV Aleksandr V.[2]**

## Introduction

The problem of gender segregation in the labor sphere is debatable, multifaceted and remains unresolved as of the present day, affecting the socio-demographic, economic, moral and ethical aspects of social life. For more than eighty years, periodically, with varying force, the gender equality questions have drawn the attention of both domestic and world science, there are appearing the views about the need to develop the new and further improve the already implemented methods of evaluation of the gender proportionality in the sphere of labor with the purpose of development or correction of the measures aimed at achieving the gender equality. This is confirmed in the 2019 UN Human Development Report: “Gender imbalances are one of the most entrenched forms of inequality across the board. Since these adverse conditions affect half of the world's population, gender inequality is one of the greatest obstacles to human development. ” [UN Report 2019]

The gender equality is assigned a significant role in the modern world, which is based upon the transformations taking place in public life within the context of humanitarian and technological revolution. The changes taking place in the sphere of labor are not gender-neutral and are reflected in the distribution of roles in the family, the models of family, and demographics, which includes the reproductive behavior. With a well-implemented policy, the change of the paradigm of socio-economic development from “people for the economy” to “economy for people” should contribute to the “gender transition”, and, as a consequence, overcoming the obstacles to the demographic shifts and growth in the level and quality of life of the population.

[1] This research was performed in the framework of the state task in the field of scientific activity of the Ministry of Science and Higher Education of the Russian Federation, project "Development of the methodology and a software platform for the construction of digital twins, intellectual analysis and forecast of complex economic systems", grant no. FSSW-2020-0008.

[1] Cand. Sci. (Economics), Associate professor of the Department of statistics, Plekhanov Russian University of Economics, Moscow, Russia. E-mail: zolotareva.oa@rea.ru

[2] Cand. Sci. (Economics), Associate professor of the Department of statistics, Plekhanov Russian University of Economics, Moscow, Russia. E-mail: bezrukov.av@rea.ru

The concept of “gender transition” itself suggests certain qualitative changes in the demographic development under the influence of the gender equality factor [Kalabikhina, 2009] that is realistically (and not formally) is achieved at the final stage of this transition in all spheres of social life, firstly while creating the gender-equal possibilities in the sphere of labor (which is, we should note, is substantially promoted by distance employment) and the elimination of wage differentiation between men and women. Let us mark that the discrimination towards women is recognized as a human rights issue on the international level. In 1945, the Charter of the United Nations (UN) first affirmed the principle of equality between women and men. [UN: Gender Dimensions ...] Through subsequent international treaties - the International Covenant on Economic, Social and Cultural Rights (1966); The Convention on the Elimination of All Forms of Discrimination against Women (1979); Platforms for Action and the Beijing Declaration, born of the 4th Beijing World Conference on Women (1995) and others - the scope of the human rights field with regard to gender equality issues has been greatly expanded.

The issues of ensuring gender equality were raised on the sidelines of the 70th anniversary session of the UN Summit, held in New York on September 25, 2015, within the framework of which the final document “Transforming our world: the 2030 Agenda for Sustainable Development” was adopted. "Among the Sustainable Development Goals (hereinafter - the SDGs), presented in the" Agenda ", the following is highlighted:" Ensuring gender equality and empowering women and girls. " [Transforming Our World: An Agenda ...]

The SDGs "gender equality" are reflected in the "National strategy of action for women for 2017-2022", approved by the order of the Government of the Russian Federation dated March 8, 2017 N 410-r. According to this Strategy, "by 2022, a system of measures should be formed to ensure the implementation of the principle of equal rights and freedoms for men and women and the creation of equal opportunities for their implementation by women in all spheres of life" [Order of the Government of the Russian Federation of 03/08/2017 N 410-r ...]

The implementation of the “gender transition” process [Kalabikhina, 2009] should contribute to the fulfillment of reproductive attitudes that enable to achieve not only the growth of birth rates, but the level of simple reproduction in the country, and achieve a higher well-being level of both families and the society as a whole, which is, undoubtedly, a life quality growth factor and an overall driver of socio-economic development. This requires, however, the development of measures that allow to carry out this transition in full, which is only possible with qualitative analysis of changes in parameters that characterize socio-economic processes in the gender aspect. Accounting the above-stated, we consider that close attention is mandatory to obtaining gender assessments in the sphere of labor, that would allow to characterize the effectiveness of social policy aimed at achieving gender equality.

## Gender analytics in the context of strategic social and demographic goals

Ensuring the national security of the state and its sustainable development, are both linked with its demographic security. In the recent decades, a special significance is taken by such demographic threats as population aging (being the consequence of labor force aging), the lack of simple reproduction of the population (the consequence of lack of natural replenishment of labor force), demographic expansion (the consequence of labor market deformation by types of economic activity, when most jobs are filled with migrants and in certain extreme cases the expansion when in a particular territorial unit a large proportion of the employed are migrants)

Today, the strategic goals approved in the Decree of the President of the Russian Federation of May 7, 2018 No. 204 "On national goals and strategic objectives of the development of the Russian Federation for the period up to 2024" have been adjusted, in which demography issues were reflected in two main directions: firstly, to the Government The Russian Federation was charged with ensuring the achievement of sustainable natural growth in the population of the Russian Federation, and secondly, an increase in life expectancy to 78 years (by 2030 - to 80 years). Note that among the national goals were those that are aimed at solving socio-demographic problems: ensuring sustainable growth in real incomes of citizens, as well as an increase in the level of pension provision above the inflation rate; halving the poverty level in the Russian Federation; improving the living conditions of at least 5 million families annually. [Decree of the President of the Russian Federation of May 7, 2018 No. 204 ...]

The adjustment / updating of national goals, taking into account the current situation (a difficult situation, primarily due to the spread of Covid-19, as well as a number of external shocks, including the continuing anti-Russian position of the West) retained a humanitarian and technological focus, while achieving updated target values of indicators seems more likely. In accordance with the Decree of the President of the Russian Federation of July 21, 2020 N 474 "On the national development goals of the Russian Federation for the period up to 2030", the Government of the Russian Federation is charged with ensuring sustainable population growth (the emphasis on natural growth is "no longer valid"). [Decree of the President of the Russian Federation of July 21, 2020 N 474 ...]

This adjustment was required, in our opinion, regardless of the pandemic, which is evidenced by the demographic situation: a smaller cohort that is in active reproductive age at the present time, formed on the basis of those born in the 1990s when there was a low birth rate, will be unable to provide extensive reproduction under the existing two-child family model; the mortality rate given significant investments in the development of medicine will not decline significantly in the coming years, but will most likely be increasing, as there are higher levels of aging in the population and the probability of dying is, undoubtedly, significantly greater in older ages and therefore a larger cohort will pass away.

As of January 1, 2020, the population of Russia, according to Rosstat, was 146.7 million people, an increase of 1.5 million people since the 2002 All-Russian Population Census (it is necessary to realize that in March 2014 Russia annexed The Republic of Crimea and the city of Sevastopol with the population of these territories).

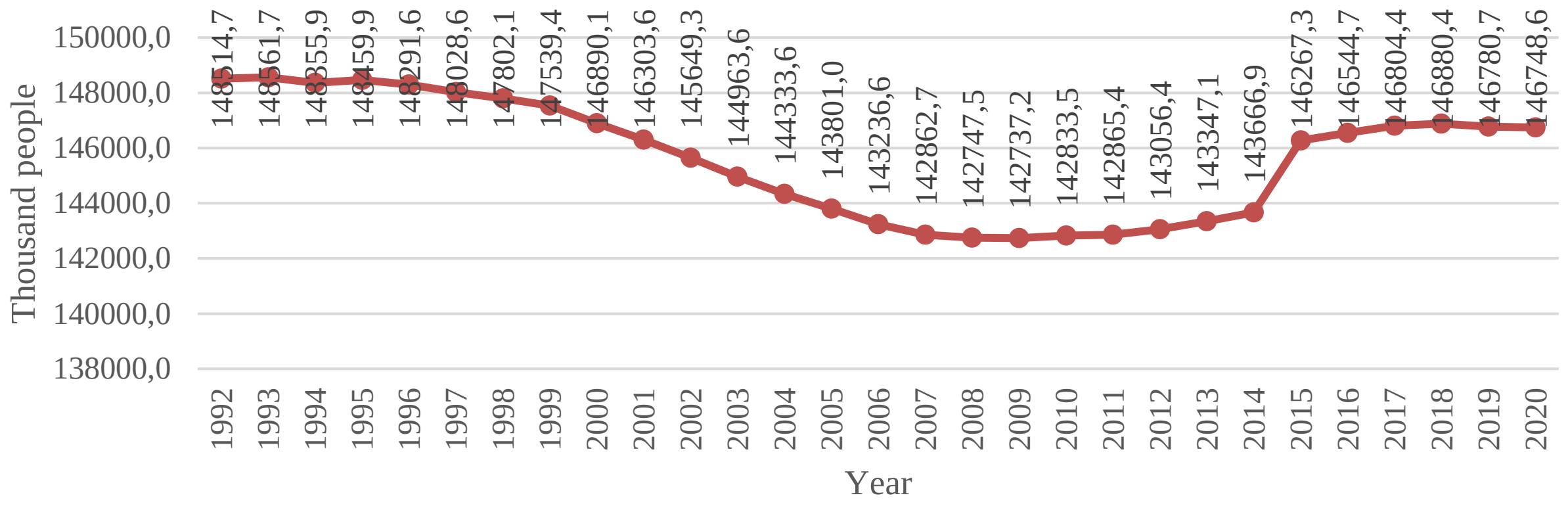

**Figure 1 - Dynamics of the population of the Russian Federation as of January 1, 1992 –2020, thousand people**

In 2020, as compared to 2019 (as of January 1) a decrease in our country’s population was recorded. These days we mark further decline in population size and the second depopulation wave, due to, as mentioned earlier, due to aging population and narrowed reproduction. Historically, the regime of narrowed population reproduction in Russia was formed in 1960s and became a certain “demographic setting” that could not be changed thus far. Over the period 1992 to 2019, the net reproduction ratio decreased from 0.732 to 0.719, while over the last year its value decreased by 4.39% (in 2018, the net reproduction ratio was 0.752).

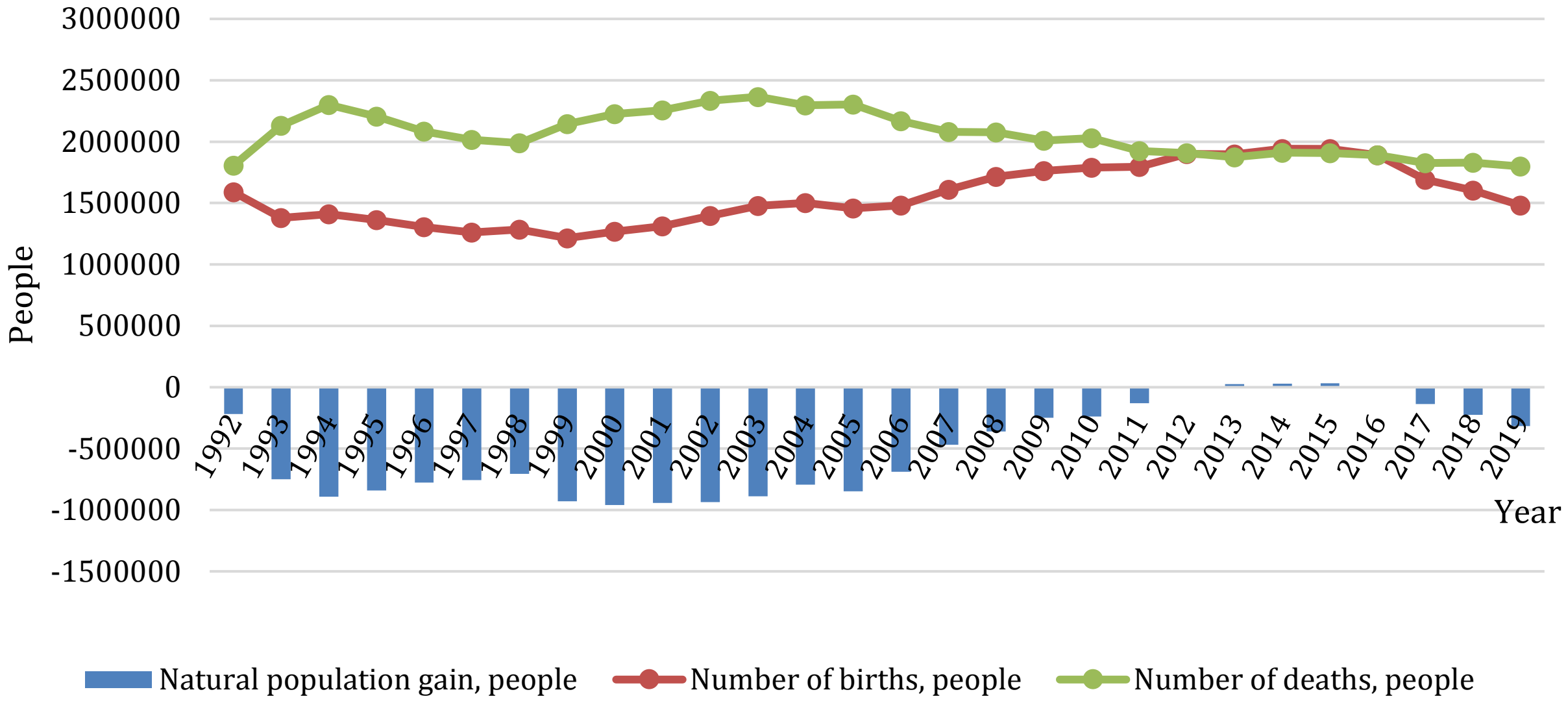

**Figure 2 - Dynamics of the number of births and deaths in Russian Federation for the period from 1992 to 2019, people**

The maximum natural population reduction from 1992 to the present was typical for 2000 and amounted to -958,5 thousand people. From the 2002 census to 2015 in the Russian Federation there was recorded a decrease in the natural population decline by 74,41 thousand people annually, which in 2015 made it possible to achieve a natural increase of 32.04 thousand people. (0,3 ‰). The number of births in 2002 to 2014 increased annually on average by 2,79%, while the average annual mortality reduction rate was -1,64%, which enabled to approach in 2013-2015 the target levels of national development for sustainable population growth. To a greater extent, the natality growth was due to the “Demographic echo” of the second half of the 1980s on the wave of expectation of positive changes (as this generation was entering the childbearing age). A significant growth of the overall births number from 1999 to 2015 was largely due to the numerous generation of girls born in 1980s by the “baby-boomers” of the 1950s and early 1960s, that entered fertile age in the 2000s. However, already in 2015-2019 the number of births started to decrease annually by 92.85 thousand people (by 5.31%). This has again been caused by the “demographic echo”, as the generation of 1980s was being replaced by the generation of 1990s which was approximately half in size, comparatively. The rate of replacement of women of reproductive age for the next 5 years will be 75,69 %, that is over the period of 2020 to 2024 per every 1000 women leaving the reproductive age (that is entering the age class of 50 and older) there will be an average of 757 girls entering reproductive age (that is entering the age of 15 and older). At the same time the proportion of girls aged 10-14 now is 4.85%, while the proportion of girls aged 15-19 is 4.31%, the women aged 20-24 are 4.43%, the women aged 25-29 years are 6.36%, and the women aged 30-34 are 8.05% from the total number of women.

In 2019 the modal maternal age was 30 years by the total number of births (the women aged 30 gave the most births, amounting to 100406 people), 25 years by the first born child (40688 first children born to women aged 25). Thus the active reproductive age will be entered by a smaller cohort in the near future, than the one leaving this age. A smaller number of women women given the reproduction level and the onset of the crisis associated with the pandemic (that entails the increase in unemployment, lower wages, higher prices etc.), will be unable, in our opinion, to give birth to a greater number of children than their predecessors.

All formed preconditions do not allow to declare a steady natural population increase, and moreover, there is an understanding that in the foreseeable future there will be a natural decline in the population. The above-stated allows to mark the weakness of analytics in public administration organs, that visualizes the real demographic picture, but this hasn’t been done at the stage of preparing the decision on providing steady natural population increase. Nevertheless, albeit with an emphasis on the pandemic. the national targets adjustments have been made, but, in our opinion, such errors should not have been made initially. As hardly achieveable there also seems the increase in the total fertility rate by 2024 to 1,7 children per woman, that is stated as a target indicator in the passport of the national “Demography” project developed by the Ministry of Labor of Russia and approved by the decision of the Presidium of the Council under the President of the Russian Federation for Strategic Development and national projects on December 24, 2018. [National project "Demography" ...] Note that the national project "Demography" was adopted pursuant to the Decree of the President of the Russian Federation dated May 7, 2018 No. 204. Currently, it needs updating.

The typical values of the total fertility rate for the recent years are shown in Figure 3. According to Rosstat, the indicator in 2019 was 1.504 children per woman on average - the goal (1.63) was not achieved - the percentage of implementation of the plan was 92.27%.

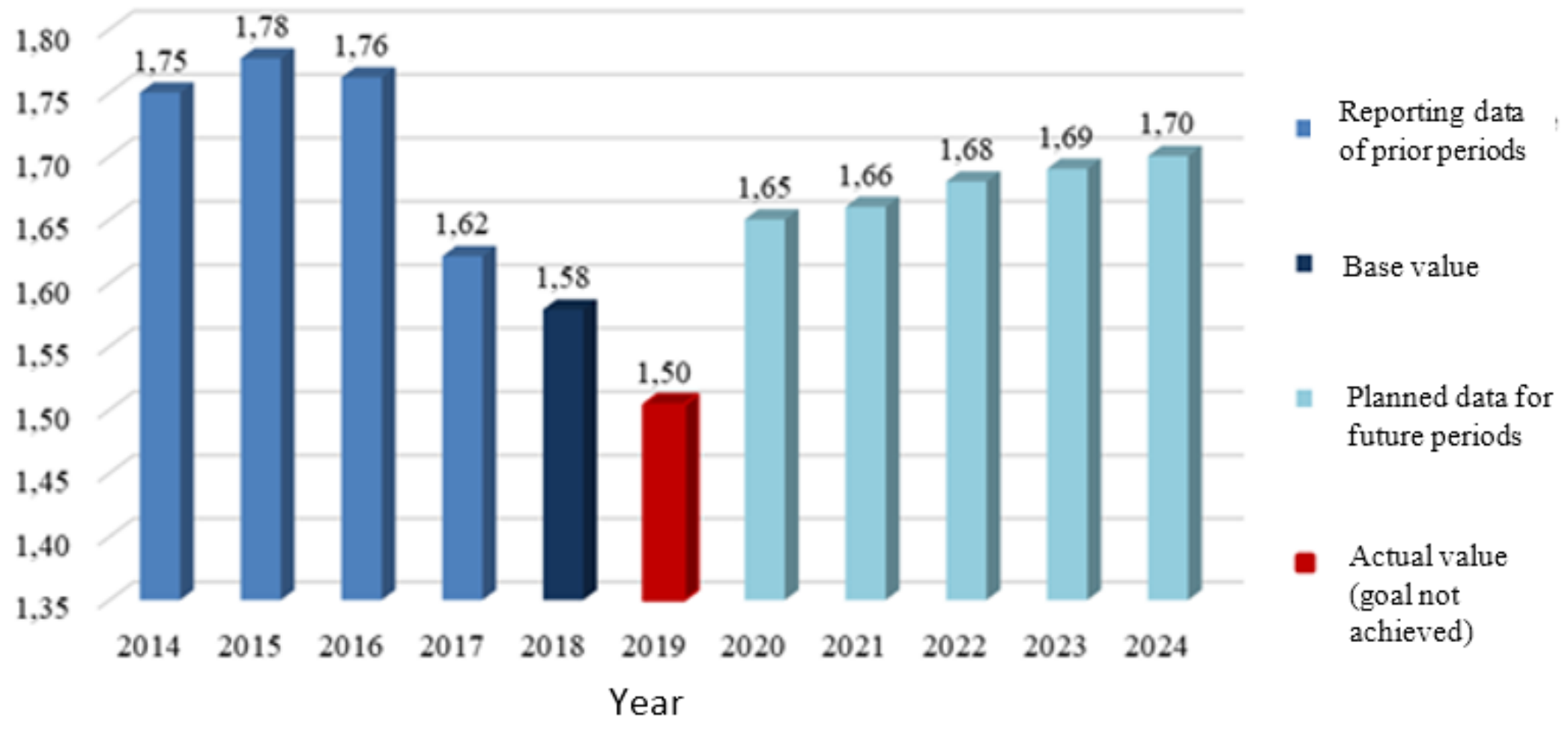

**Figure 3 - Dynamics of the total fertility rate in the Russian Federation, actual values for the period from 2014 to 2019 and planned values in accordance with the national project for the period from 2020 to 2024, (number of children per 1 woman)**

The presented data allow to state that in Russia a two-child family model has been formed, and a transition to a multi-child family seems to be quite difficult in the modern conditions, and, we can also say, an impossible task in the coming years. This is confirmed by the results of the all-Russian sociological survey “Demographic well-being of Russia” conducted by ISPI RAS‡, according to which ideas about childhood in the family have been revealed: in general, society has adopted a two-child family model.

‡ All-Russian sociological survey "Demographic well-being of Russia", conducted at the end of 2019 - beginning of 2020. in the Central, North-West, Volga, Ural, North Caucasian, Southern Federal Districts. N = 5616 representatives of different generations aged 18 to 50 years. Hands. Doctor of Social Sciences, Professor T.K. Rostovskaya.

Among the respondents (aged 30-39 years and 40-50 years) with children, and among the opinions on the desired number of children, two children are an optimal / preferrable option (either as a fait accompli or as an assessment for the future). At the same time, the family model transformation is uninfluenced by the presence of all necessary conditions for the birth of children and their upbringing, and it remains a two-children model. Moreover, among the main factors preventing them from having more children than they intend to, women indicate earnings difficulties (63.1% of respondents), uncertainty about the future (63.0%). This states of the existing problems for women, primarily associated with the ability to harmoniously combine the birth and upbringing of a child with participating in labor activity. The digitalization in overall should help to overcome these difficulties, since it enables remote work, but here it is also hardly possible to speak only of positive effects. The transformation of the labor sphere within the context of active introduction of technologies leads to the release of job vacancies, and the disappearance of an array of "traditional professions". There are observed and will intensify the structural changes in employment. The workforce is being redistributed in the context of new requirements of the digital economy, but at the same time there is an acute shortage of highly qualified personnel. The gender problem is also intensifying, as the main barrier that prevents women from being fully included in the digital economy, is their low professional training as highly qualified specialists in the field of STEM (science, technology, engineering and mathematics). It should be noted that in STEM-employment and in STEM-education there have formed certain gender stereotypes which are quite difficult to "break".

A relevant target indicator that reflects the achievement of the national goal of "Preserving the population, health and well-being of people" – the increase of life expectancy to 78 years to 2030 – does not imply a gender component. Over the past few decades, the differences between the life expectancy of men and women were above 10 years since 1989, and only in 2019 this lowered to a slightly lesser value of 9.93 years. In the world, the gender differences in this characteristic are normal and reasonable. This suggests the need to take them into account, especially within the modern conditions (how the consequences of the pandemic will be gender-related).

The two previously outlined goals, namely ensuring sustainable growth in real incomes of citizens, and the increase in the level of pension provision above the inflation rate, and halving the level of poverty in the Russian Federation - at the time of the signing of the decree, seemed rather difficult to implement, not to mention the present state of the situation. An Academician of the Russian Academy of Sciences Abel Gezevich Aganbegyan in his report at the 26th expert session of the VEO Coordination Club of Russia pointed out that “over the past seven years of stagnation (2013-2019) during the recession in 2015, there was a noticeable reduction: real incomes of the population by 10, 4%; retail turnover per capita by 9.3%; final consumption of households by 1.5% ”. [Aganbegyan A.G., 2020] Moreover, the crisis provoked / associated with a pandemic primarily affected the service sector, in which, according to Rosstat, in 2019 more than 60% were employed [Rosstat. Labor market ...] of all employed in the economy in connection, which determines a massive increase in the number of unemployed. It will be possible to partially avoid a significant increase in the unemployment rate with the redistribution of the employed in the context of natural / inevitable structural changes in the economy, which, as noted earlier, are taking place, associated with digitalization. Presumably, in this aspect, the consequences can be characterized in two ways, on the one hand, positively - an increase in the spread of distance learning and employment was recorded, that is, the period of the pandemic was a kind of "impetus" for the development of digitalization, technologies were rapidly introduced into the economy and social sphere, on the other - negatively - empowerment, inequality in society is increasing, as stated in the UN Human Development Report 2019 [UN Report 2019]. Problems in the world of work, which were also observed before the digital age, such as gender inequality in employment, and wage differentiation will only get stricter. Moreover, in the conditions of "holes" in the federal and regional budgets, which are formed due to a decrease in tax revenues (first of all, from taxes on corporate profits and on personal income, customs duties) and the implementation of support measures today (and for population, they are short-term and are quickly exhausted), as well as a decrease in the total amount of the Pension Fund, there is no need to talk about a halving of the poverty level in the country today. At the same time, in the Decree of the President of the Russian Federation of July 21, 2020 No. 474 “On the national development goals of the Russian Federation for the period up to 2030” as a target indicator characterizing the achievement of strategic goals, an indicator is given: “the poverty level is halved in comparison with the indicator of 2017 ". [Decree of the President of the Russian Federation of July 21, 2020 N 474 ...]

The evaluation of the real picture of gender disproportionality / asymmetry in the sphere of labor and in the salary has marked the presence of a problem.

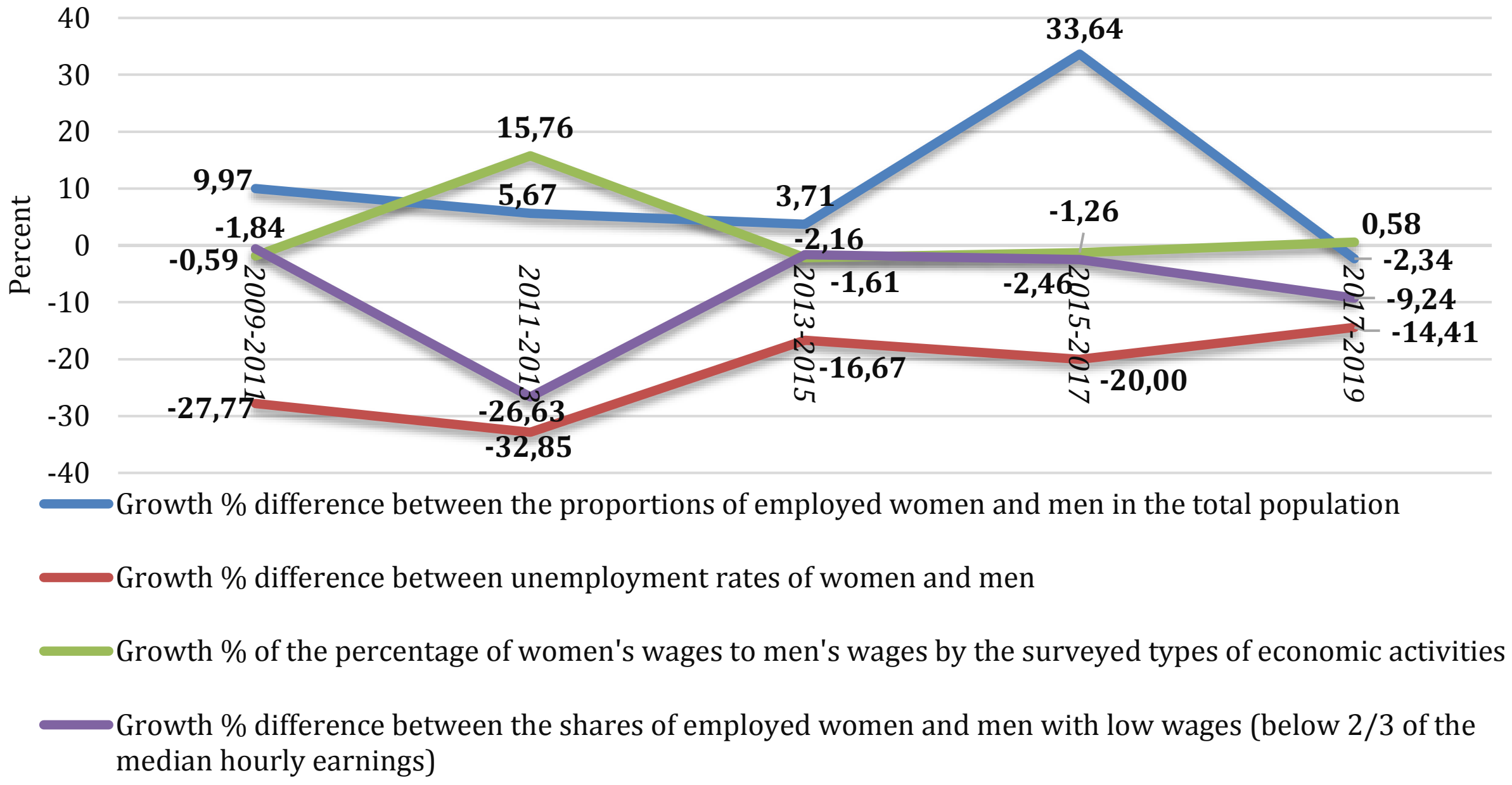

**Рисунок 4 – Темпы прироста основных показателей, характеризующих гендерные диспропорции в сфере труда в России за период 2009-2019гг., %**

The gender disproportionality in the level of employment in 2019 compared to 2009 has increased. The absolute difference between the separate proportions of the employed men and women in the overall number of corresponding gender grou[s increased, over the studied period, from 9.13 percentage points (p.p.) to 14.36 p.p. In 2019, the proportion of employment was 52.9% among women and 67.3% among men.

The gender differences are more significantly indicated in employment by vocation, that are perspective in the context of Society 5.0 development.

Rosstat provides the data on sample surveys of the population by the problems of employment (the survey of labor force)[§], that indicate a somewhat low employment of women in the STEM sphere and substantial gender skewness towards men.

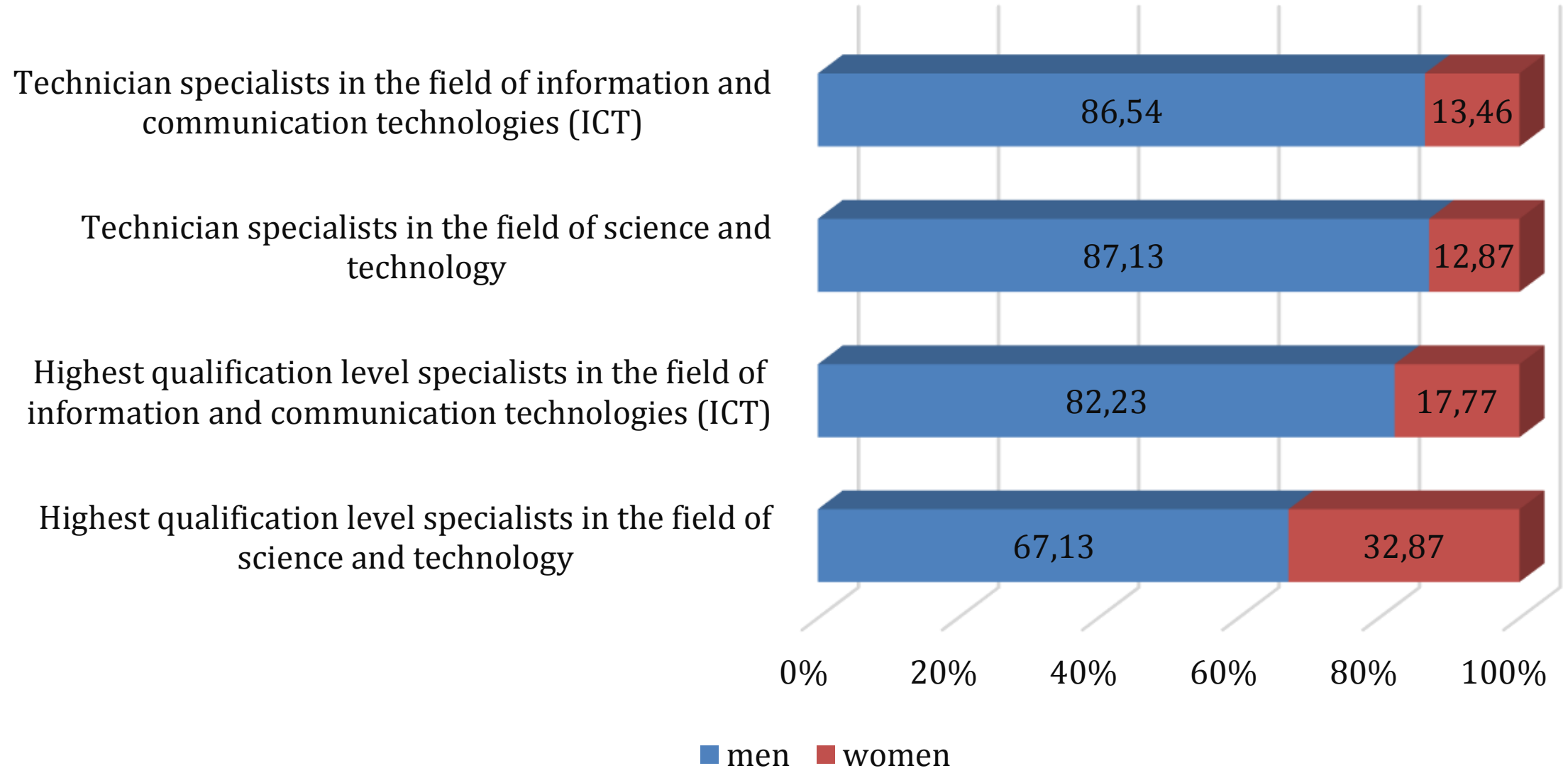

**Figure 5 - Gender structure of the employed population aged 15 and over by a number of occupations at the main job in Russia in the 1st quarter of 2020,% (based on the results of sample labor force surveys)**

In the first quarter of 2020 among the specialists with the highest qualification level in the area of science and technology the proportion of women is 32.87%, and 12.87% with the mid-level of qualification, while the number of female specialists with the highest qualification in the sphere of science and technology increased by 8.31% over the one-year period (it was 962 th.people in the Ist quarter of 2019 and 1042 th.people in the Ist quarter of 2020). Among the highest-qualification specialists in the sphere of information and communication technology (ICT) the proportion of women was 17.77% and the mid-qualification female specialists were 13.46%. The number of women specialists with the highest level of ICT qualifications in the first quarter of 2019 was 129 thousand people, and 205 thousand people in the first quarter of 2020 (an increase of 58.78%). This confirms the traditionality of the STEM sphere as "male" in terms of activities and professions, although within them the physical condition of a person does not appear to be a determining factor.

---

§ Росстат. Обследование рабочей силы. Электронный ресурс: http://www.gks.ru/wps/wcm/connect/rosstat_main/rosstat/ru/statistics/publications/catalog/doc_1140097038766

The gender differences are indicated not only in the employment but in the salary of workers as well, and we can note that in overall there is a certain shrinkage of the gap between the men and women salaries (Fig.4), although it remains significant. The average accrued wages of women and men for the surveyed types of economic activity (upon the results of sample surveys of organizations in October) in 2019 amounted to 37872 rubles and 52533 rubles, respectively. The ratio of women's salary to men's salary for the surveyed types of economic activity was 65.3% in 2009 and reduced to 72.1% by 2019.

The Rosstat presents additional indicators that characterize equal opportunities at work, that were developed in correspondence with recommendations of the International Labor Organizations and are part of the system of decent labor indicators. One such indicator is the gender wage gap, that is determined by the following formula: (average hourly wages for men / average hourly wages for women) x 100% , excluding financial organizations; public administration and military security; social insurance; delivery of public services and activities of public organizations.[**]

The gender gap in wages of workers in 2009 was 29.2% (Table 1), and reduced to 24.8% by 2019, which confirms the above analysis of the ratio between the average accrued wages for men and women. In 2018 the proportion of employed women with low wages (below 2/3 of the median hourly earnings) was 10.8 percentage points higher than the proportion of employed men with low wages.

---

[**] Decent Labor Indicators. Methodological explanations for the calculation. URL: http://www.gks.ru/wps/wcm/connect/rosstat_main/rosstat/ru/statistics/wages/

**Table 1 - Dynamics of indicators characterizing equal / unequal gender opportunities and relationships at work, from those recommended by the International Labor Organization** [††]

| Indicator | 2009 | 2011 | 2013 | 2015 | 2017 | 2019 | Source |
|---|---|---|---|---|---|---|---|
| Gender wage gap,% | 29,2 | 29,0 | 22,8 | 24,4 | 25,3 | 24,8 | Rosstat. Since 2005 - sample surveys of organizations (once every 2 years). |
| Share of employed men with low wages (below 2/3 of the median hourly earnings), % | 19,5 | 19,6 | 21,4 | 20,7 | 20,0 | 19,0 | Rosstat. Sample surveys of organizations (excluding small businesses), for October, are carried out once every two years. Until 2009, the indicator was not calculated. |
| Share of employed women with low wages (below 2/3 of the median hourly earnings), % | 36,5 | 36,5 | 33,8 | 32,9 | 31,9 | 29,8 | |

It is necessary to consider the gender disparities in the average accrued wages by types of economic activity with the purpose of carrying out the effective narrowly focused socio-economic policy. Only taking into account the need to expose and eliminate the disproportions in the price of male and female labor that have developed in specific types of economic activity, it would be possible to achieve gender equality in the economy as a whole.

[††] Decent Labor Indicators. URL: http://www.gks.ru/wps/wcm/connect/rosstat_main/rosstat/ru/statistics/wages/

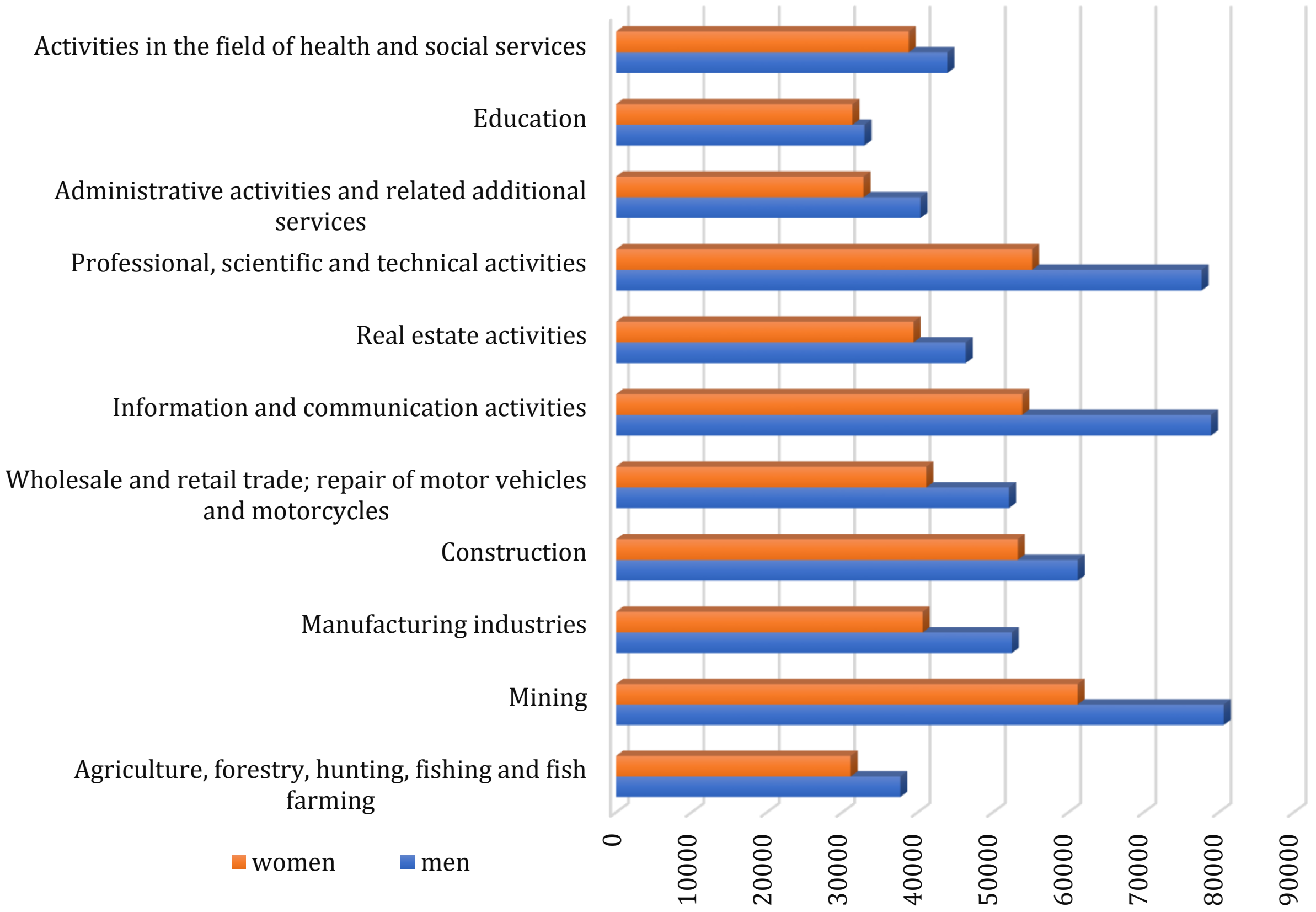

**Figure 6 - Average accrued wages of men and women for a number of surveyed types of economic activity in Russia in 2019, rubles (based on the results of sample surveys of organizations in October)**

In October 2019 the most significant gender disparities were recorded in the average accrued wages for the type of economic activity of field of information and communication, which was 78 980 rubles for men and 53 887 rubles for women; and the ratio of women's wages to men's wages was 68,23%. The gender difference in the average accrued wages is the highest among the types of economic activities and is equal to 25 093 rubles. Additionally, highly substantial gender disparities were observed in the average accrued wage in the sphere of professional, scientific and technological activity, where the ratio of women's wages to men's wages was 71,01% (the average men's wage was 77 719 rubles, and the average women's wage was 55 187 rubles). The average accrued wage was only higher for the workers employed in the economic activity of "Mining", amounting to 80 663 rubles for men, and 61 246 rubles for women (75,93% of the average accrued wages of men).

This, for the high-paying types of economic activity a more substantial gender gap is typical, that will be further increased during digitalization. The more in demand and higher-paid specialists are the ones involved in the development of new production and administration technologies based upon digital solutions. In the sphere of administration and in the sphere of ITC alike the steady gender skewness has developed.

The existing gender segregation in the sphere of labor (including the gender differentiation in salary that is especially typical for the IT sphere) is a barrier in the area of demographic security. Nowadays, the successful combination of work and family life is the priority driver of increase in natality, but this requires the achieving of gender equality in the sphere of labor and the elimination of traditional cultural barriers.

**CONCLUSION**

Inarguably, at present the state carries out an active policy aimed at the growth of economy and Russian population life quality. However, the performed analysis indicated insufficient analytical provision of strategic planning / administration, that is associated with low level of analytical literacy. The socio-demographic processes are characterized by dynamicity, multifactoriality and are subject to significant risks in their development in the context of external shocks, uncertainty of situation changes in the country and in the world, and in this regard there is needed the effective administration that implies quality targeting, forecasting and planning, based upon reliable and relevant information that characterizes their state and their changes. Today, there is needed the quality strategic planning / administration, while the strategic documents must not contain inherently unachievable goals. The performed evaluation indicated that within the context of developed demographic situation and the recorded demographic waves, in the nearest perspective a natural population reduction will be observed, that is suggested to be alleviated by migration. However, in our opinion, considering migration as “the cure” or a certain “patch” that would allow to “stitch the edges of the demographic pit” is incorrect. Moreover, the increase of average life of the Russians to 78 years with simultaneous reduction of poverty appear to be considerably hard goals to achieve, that, under the present conditions, we can deem as impossible.

The existing crisis, indisputably, introduces its corrections in the already set goals / plans, and, to much regret, almost never towards positive. The problems considered in the article are exacerbated, as the differentiation in population income is increasing while the poverty is increasing as well. It is worth noting, though, that the unfavorable epidemiological situation (the pandemic) has become a contributor to the remote employment and remote learning, which allows to speak of a possible breakthrough in the achievement of the goal of ensuring the accelerated introduction of digital technologies in the economy and the social sphere. However, despite the existing breakthroughs under the influence of external shocks in the formation of digital Society 5.0, it is important to adequately assess the situation, with, firstly, understanding the changes in the economy structure will cause a free surplus of labor force. Besides this problem a number of obstacles is additionally revealed for effective implementation of digitalization, such as the “landscape” is unprepared, the deficiency of highly qualified personnel is observed, that could be used as labor resources in the framework of the digital economy. Moreover, the transformations are contained in the fact that the digital Society 5.0 reveals new opportunities for women and defines new gender challenges alike.

At present, it is necessary to account the importance of overcoming long-standing gender stereotypes about the possibility of women participating in the STEM sphere that are historically sex-role in nature, while the gender segregation remains on all levels of employment. [Khasbulatova, 2018] It is necessary to emphasise the timely and high-quality provision of vocational training and additional vocational education for women on maternity leave (until their children reach the age of three). Within the structure of the unemployed the proportion of women with higher professional education is increasing, which is associated with the loss of significance of the knowledge gained in the absence of opportunity of updating it.

In the digital age, the knowledge intensity of the economy is increasing with the rapid global renewal of knowledge itself. At the same time, we would like to draw your attention to the fact that according to the additional and substantiating materials of the federal project "Promoting the employment of women - the availability of preschool education for children" implemented within the framework of the national project "demography", it is considered sufficient that the state fulfills its obligations to provide children under three years of age with places in organizations of preschool education and the organization of retraining and advanced training of women during the period of leave to care for a child under the age of three years as factors in the growth of employment of women raising children of preschool age. [Federal project “Promotion of the employment of women ...]

The above-stated measures are significant, but by no means are sufficient, and, moreover, need more specification and clarification (regarding the measures of retraining and advanced training). The processes of retraining organization and advanced training organization require the approach that accounts the understanding of the ongoing transformation in the sphere of employment and the awareness of the importance of STEM professions. It is necessary to be “one step ahead” and try to avoid the essentially meaningless expenditures on professional education that would be partially or fully automated in the near future (such as accountants, bank teller, loan manager etc.). It is crucial to develop a complex of targeted measures that would allow, in the foreseeable future, to give the women the education meeting the modern requirements in the context of digital development, and provide them with the opportunity to master the STEM-professions and the corresponding skills and competencies. The resolution of these issues would contribute to substantial growth of the involvement of women in the STEM-economy. In their own turn, the promising STEM-professions allow remote employment, that would substantially contribute to expanding the opportunities of full-fledged self-realization of women both as mothers and keepers of the home, and as highly qualified specialists.

Under the conditions of the changing paradigm of socio-economic development in the context of humanitarian and technological revolution the implementation of "gender transition" is possible. Only by achieving the true gender equality that allows women to feel more confident in the modern society, it is possible to reverse the current trend of loaned reproduction and achieve sustainable demographic development.

In overall, the adjustment and updating of national goals shows that certain work has been done, but there is still more to be resolved, for instance, to update / develop additional (if necessary) national projects, but this work, however, must be carried out with the account of expert evaluations and proposals from the scientific community.